\documentclass[twocolumn,showpacs,preprintnumbers,amsmath,amssymb]{revtex4}

\usepackage{graphicx}
\usepackage{dcolumn}
\usepackage{bm}
\usepackage{bbm}
\usepackage{placeins}

\newcommand\Id{\ensuremath{\mathbbm{1}}}

\begin{document}

\preprint{ }

\title{Gamow shell model description of proton scattering on $^{18}$Ne}

\author{Y. Jaganathen}
\affiliation{Department of Physics and Astronomy, University of Tennessee, Knoxville, Tennessee 37996, USA}
\affiliation{Joint Institute of Nuclear Physics and Applications, Oak Ridge National Laboratory, Oak Ridge, Tennessee 37831, USA}

\author{N. Michel}
\affiliation{
Grand Acc\'el\'erateur National d'Ions Lourds (GANIL), CEA/DSM - CNRS/IN2P3,
BP 55027, F-14076 Caen Cedex, France
}%

\author{M. P{\l}oszajczak}
\affiliation{
Grand Acc\'el\'erateur National d'Ions Lourds (GANIL), CEA/DSM - CNRS/IN2P3,
BP 55027, F-14076 Caen Cedex, France
}%

\date{\today}

\begin{abstract}
\begin{description} 
\item[Background:] Structure of weakly bound/unbound nuclei close to particle drip lines is different from that around the valley of beta stability. A comprehensive description of these systems  goes beyond standard shell model (SM) and demands an open quantum system description of the nuclear many-body system. 
\item[Purpose:] For that purpose, we are using the Gamow shell model (GSM) which provides a fully microscopic description of bound and unbound nuclear states, nuclear decays, and reactions. We formulate the GSM in coupled-channel (GSM-CC) representation to describe low-energy elastic and inelastic scattering of protons on $^{18}$Ne. 
\item[Method:]  The GSM-CC formalism is applied to a translationally-invariant Hamiltonian with an effective finite-range two-body interaction. We discuss in details the GSM-CC formalism in coordinate space and give the description of the novel equivalent potential method for solving the GSM-CC system of integro-differential equations. This method is then applied for the description of $(p,p')$ reaction cross-sections. Reactions channels are built by GSM wave functions for the ground state $0^+$ and the first excited $2^+$ of $^{18}$Ne and a proton wave function expanded in different partial waves. The completeness of this basis is verified by comparing GSM and GSM-CC energies of low-energy resonant states in $^{19}$Na. Differences between the two calculations provide a measure of missing configurations in the GSM-CC calculation of low-energy states of $^{19}$Na due to the restriction on the number of excited states of $^{18}$Ne.
\item[Results:] We present the first application of the GSM-CC formalism for the calculation of excited states of $^{18}$Ne and $^{19}$Na, excitation function and the elastic/inelastic differential cross-sections in the  $^{18}$Ne$(p,p')$ reaction at different energies. This is the first unified description of the spectra and reaction cross-sections in the GSM formalism. The method is shown to be both feasible and accurate. The approximate equivalence of GSM and GSM-CC in describing spectra of $^{19}$Na has been demonstrated numerically. 
\item[Conclusions:] The GSM in the coupled-channel representation opens a possibility for the unified description of low-energy nuclear structure and reactions using the same Hamiltonian. While both GSM and GSM-CC can describe energies, widths and wave functions of the many-body states, the GSM-CC can in addition yield reaction cross-sections. Combined application of GSM and GSM-CC to describe energies of resonant states allows to test the exactitude of calculated cross-sections for a given many-body Hamiltonian.
\end{description}
\end{abstract}

\pacs{24.10.-i, 24.10.Cn, 24.50.+g, 21.10.-k}

\bigskip

\maketitle

\section{Introduction}

The unified description of both structure and reactions of nuclei in terms of the constituent interacting nucleons is the long-term goal of nuclear theory. The first attempts to reconcile the SM with the reaction  theory go back to Feshbach and his projection formalism\cite{fesh} which inspired the continuum shell model (CSM)\cite{Mah69} and evolved into a unified of theory of structural properties and reactions\cite{Mah69,Bar77,Ben00,Vol06} with up to two nucleons in the scattering continuum\cite{Vol05}. 

{\em Ab initio} description of bound states of light nuclei became possible in terms of realistic nucleon-nucleon and three-nucleon interactions\cite{Nav00,Pie01,Bar04,Hag10,Pap13}. First {\em ab initio} scattering calculations were performed using the Green's Function Monte-Carlo method\cite{Nol07} with two- and three-body interactions.  More recently, the {\em ab initio} approach to low-energy reactions\cite{Nav08} has been proposed by combining the resonating-group method (RGM)\cite{rgm} and the no-core shell model (NCSM)\cite{Nav00}. In this approach, one assumes that nucleons are grouped in clusters. The RGM provides then the correct asymptotic of the multi-cluster wave function, whereas each cluster wave function is described using the microscopic NCSM wave function, neglecting the continuum coupling. Up to now, applications of the NCSM/RGM approach were based on the binary- and ternary-cluster wave functions\cite{Nav08,Nav11,Nav13} with two- and three-body realistic interactions\cite{Nav_arch}. 

A most general treatment of couplings between discrete and scattering states is possible in the framework of GSM\cite{GSM,GSM1,GSM2}. In the GSM, a single-particle (s.p.) basis is given by the Berggren ensemble\cite{berggren} which consists of Gamow (resonant) states and the non-resonant continuum.  The GSM Hamiltonian is Hermitian. However, since the s.p. vectors  have either outgoing or scattering  asymptotics,  the Hamiltonian matrix in GSM is complex symmetric and its eigenvalues are complex above the first particle emission threshold. The GSM offers a fully symmetric treatment of bound, resonance, and scattering s.p. states and contains all salient features of an interplay between opposite effects of Hermitian and anti-Hermitian couplings. It is also a generalization of the standard nuclear SM to describe well bound, weakly bound and unbound many-body states. Another {\em ab initio} approach which applies the Berggren ensemble is the Coupled Cluster (CC) approach\cite{hag1,hag2} which has been applied recently for the calculation of phase shifts and elastic cross-section for the scattering of protons on $^{40}$Ca target at low energies\cite{Hagmic}.

So far, GSM has been used  mainly in the context of nuclear structure.  (For a recent review, see Ref. \cite{Mic09}.) 
In this paper, we shall extend GSM  to reaction problems using a coupled-channel (CC) formulation of the scattering process. The application of the GSM-CC formalism will be presented in this paper for the proton scattering on $^{18}$Ne target. The proposed GSM-CC approach can be easily generalized for the description of nuclear reactions in the {\it ab initio} framework of the No-Core Gamow Shell Model\cite{papa12} and to heavier projectiles like deuteron or $\alpha$-particle\cite{future}. 

The paper is organised as follows. Section \ref{sec2} presents the formalism of GSM-CC approach. In Sect. \ref{sec2a} we introduce the translationally-invariant GSM Hamiltonian in the Cluster-Orbital Shell Model (COSM) variables\cite{Ikeda}. The GSM reaction wave functions and the CC equations are discussed in Sect. \ref{sec2b}. The derivation of the GSM Hamiltonian and overlap kernels for the problem of scattering of a nucleon on the many-body target is discussed in Sect. \ref{cc_eqs_derivation}.

Section \ref{sec3} presents the resolution of GSM-CC equations. A novel equivalent potential method for solving GSM-CC integro-differential equations is discussed in Sect. \ref{eq_pot_GSM}. Boundary conditions and the choice of the basis functions are explained in Sect. \ref{basis_functions}, and practical aspects of solving the CC equations are discussed in Sect.~\ref{sec3c}. 

Discussion of the cross-sections and excitation functions for the reaction $p+^{18}$Ne at various bombarding energies are contained in Section \ref{results}. The potential of the $^{16}$O core and the effective two-body interaction are presented in Sects. \ref{potential} and \ref{interaction}, respectively. $^{18}$Ne$(p,p')$ reaction cross-sections and comparisons to experimental data are discussed in Sect. \ref{detres}.

 Finally, conclusions of this work are summarized in Section \ref{sec5}.

\section{Derivation of the CC equations in coordinate space
\label{sec2}}

\subsection{The Hamiltonian of the Gamow Shell Model\label{sec2a}}

The translationally invariant GSM Hamiltonian in intrinsic nucleon-core coordinates of the cluster-orbital shell 
model\cite{Ikeda}, can be written as:
\begin{equation}
H= \sum_{i=1}^{A_{val}}\left[ \frac{p_{i}^{2}}{2\mu_i} + U_{i} \right] + \sum_{i<j}^{A_{val}} \left[ V_{ij} 
+ \frac{1}{M_c} \bf{p}_{i}\bf{p}_{j} 
\right],
\label{GSM_Hamiltonian}
\end{equation}
where $M_c$ is the mass of the core, $\mu_i$ is the reduced mass of either the proton or neutron ($1 / \mu_i =1/m_i + 1/M_{c}$),
$U$ is  the s.p. potential describing the field of the core, $V$ is the two-body residual interaction between
valence nucleons. The last term in Eq. (\ref{GSM_Hamiltonian}) represents the recoil term. 

The particle-core interaction is a sum of nuclear and Coulomb terms: $U = U^N + U^C$. The nuclear potential $U^N$ is modelized by a Woods-Saxon (WS) field with a spin-orbit term\cite{Mic03}. The Coulomb field $U^C$ is generated by a Gaussian density of $Z_c$ core protons\cite{Mic10}. 

Similarly, the residual interaction can split into nuclear and Coulomb parts: $V = V^N + V^C$, where $V^N$  is the Modified Surface Gaussian (MSG) 
interaction\cite{Mic03}. 
$V^C$ is the two-body  Coulomb interaction which can be rewritten as: 
$$U^C_{Z_{val}-1} + \left[V^C -  U^C_{Z_{val}-1}\right]^{HO} \, ,$$  
where $U^C_{Z_{val}-1}$ carries out of the asymptotic behavior of the Coulomb interaction and thus provides an accurate treatment of the long-range physics of the Coulomb potential. 
The second term in this equation and the two-body  recoil term is expanded in the harmonic oscillator basis\cite{PRC_real_inter_2006,Mic10}. In this work, we took 9 harmonic oscillator shells with the oscillator length $b = 2$\,fm.

\subsection{N-body GSM reaction wave functions\label{sec2b}}

Nuclear reactions involving the scattering of one nucleon can be conveniently described in a CC framework. Let us consider the following $A$-body spherical scattering state of a given set of angular quantum numbers $(J^\pi _A,M_A)$ expressed in reaction channels $c$:
\begin{eqnarray}
| \Phi ^{J^\pi _A}_{M_A}\rangle = \sum_{c} \int_0^{+\infty} \frac{u_{c}^{J^\pi _A}(r)}{r}\mathcal{A} | (c,r)^{J^\pi _A}_{M_A}\rangle  ~ r^2  dr \label{10}\ ,
\end{eqnarray}
 where the channel states are defined by:
 \begin{eqnarray}
\mathcal{A}| (c,r)^{J^\pi _A}_{M_A}\rangle  =   \mathcal{A}\left[ | \Psi_{c} \rangle^{J^\pi_{c}} \otimes |r\ell_{c} j_{c} \tau_c \rangle \right]^{J^\pi _A}_{M_A} \ . \label{channel_states}
\end{eqnarray}
The reaction channel $c$ is specified by the nucleon state $|\ell_{c} j_{c} \tau_c \rangle$ coupled to the $(A-1)$-body target state $| \Psi_{c} \rangle^{J^\pi_{c}}$. Both angular momenta $j_{c} $ and $J_{c}$ are coupled to $J_A$ - with $M_A$ projection, while $\ell_c$ and $\pi_c$ make a $\pi_A$-parity for the whole system. $\tau_c$  stands for the isospin quantum number (proton or neutron).  All $A$-body wave functions are fully antisymmetrized, as emphasized by the $\mathcal{A}$ symbol. $u_{c}(r)$ denotes the radial amplitude of the $c$ channel to be determined. It is a function of the radial coordinate $r$, i.e. the relative distance between the core of the target and the projectile. 

The CC equations are then obtained from projecting the Schr{\"o}dinger equation $| \Phi ^{J^\pi _A}_{M_A}\rangle = E | \Phi ^{J^\pi _A}_{M_A}\rangle$ on a given channel $c$:
\begin{eqnarray}
\sum_{c'} \!\!\int_0^{+\infty} \!\! \left[ \mathcal{H}_{cc'}^{J^\pi_A M_A} \!  \! -\! E  \mathcal{O}_{cc'}^{J^\pi_A M_A}  \right]\!\!(r,r') \frac{u_{c'}(r')}{r'} r'^2 d r'  = 0 \label{cc_eqs_formal_radial}
\end{eqnarray}
in which both the $A$-body Hamiltonian:
\begin{equation}
\mathcal{H}_{cc'}^{J^\pi_A M_A} (r,r') = \langle \mathcal{A} (c,r)^{J^\pi _A }_{M_A} \vert H\vert \mathcal{A}(c',r')^{J^\pi _A}_{M_A} \rangle ~ ,\label{Hamiltonian_kernels} 
\end{equation}
and overlap kernels
\begin{equation}
\mathcal{O}_{cc'}^{J^\pi_A M_A} (r,r') = \langle\mathcal{A}(c,r)^{J^\pi _A}_{M_A}\vert\mathcal{A} (c',r')^{J^\pi _A}_{M_A}\rangle \label{overlap_kernels} 
\end{equation}
will be derived explicitly. The presence of the antisymmetry in the equations implies the non-orthogonality of different channels rendering  
Eq. (\ref{cc_eqs_formal_radial}) a generalized eigenvalue problem.  

\medskip 

The derivation of the Hamiltonian and overlap kernels demands a more convenient formulation of the channel states.  In the GSM, the states of the target nucleus are eigenstates of the Hamiltonian and are expressed as a linear combination of $(A-1)$-body Slater determinants generated by a s.p. potential $U_{basis}$. It is natural to expand the $|r \ell_{c} j_{c} \tau_c\rangle$ nucleon state in the  s.p. basis of GSM wave functions $|n \ell_{c} j_{c} \tau_c \rangle$ generated by the same potential $U_{basis}$. 
Defining $$u_{nc}(r) = r \langle n\ell_c j_c \tau_c | r \ell_{c} j_{c} \tau_c\rangle ~ \ ,$$ one has:
\begin{eqnarray}
|r \ell_{c} j_{c} \tau_c \rangle =\sum_{n} \frac{u_{nc}(r)}{r} ~ |n \ell_{c} j_{c} \tau_c \rangle  \label{a_dagger} \ .
\end{eqnarray}
The expression of the channel wave functions becomes:
\begin{eqnarray}
\mathcal{A}| (c,r)^{J^\pi _A}_{M_A}\rangle  = \sum_{n} \frac{u_{nc}(r)}{r} \mathcal{A}| (c,n)^{J^\pi _A}_{M_A}\rangle    \ , \label{channel_state_expansion}
\end{eqnarray}
in which the evaluation of 
\begin{eqnarray}
\mathcal{A}| (c,n)^{J^\pi _A}_{M_A}\rangle = \mathcal{A}[  | \Psi_{c} \rangle^{J^\pi_{c}} \otimes |n\ell_c j_c \tau_c \rangle ]^{J^\pi _A}_{M_A} \label{coupled_Psi_rljtau}
\end{eqnarray}
 in terms of $A$-body Slater determinants is straightforward.

\subsection{Derivation of the Hamiltonian and overlap kernels} \label{cc_eqs_derivation}

In order to derive the Hamiltonian kernels of Eq. (\ref{Hamiltonian_kernels}), the Hamiltonian of Eq. (\ref{GSM_Hamiltonian}) which can be cast in the form  
\begin{eqnarray}
H=T+U_{core}+V_{res}  \label{H} \ , 
\end{eqnarray}
where $V_{res}$ and $U_{core}$  are respectively the two-body residual interaction and the potential generated by the core, is separated into basis and residual parts:
\begin{eqnarray}
H=T+U_{basis} + (V_{res}-U_0)  \label{H_Ubasis_Vres} \ .
\end{eqnarray}
 $U_{basis}$ is the optimal potential of the $A$-particle system and $U_0 = U_{basis} - U_{core}$. 
The advantage of this decomposition is that $V_{res}-U_0$ is finite-range and $T+U_{basis}$ is diagonal in the basis of Slater determinants used.

The infinite-range components of the Hamiltonian $H$, combined with the presence of the infinite sum in the channel states in Eq. (\ref{channel_state_expansion}) lead to Dirac delta's which have to be calculated analytically. 
For that purpose, we suppose that only a finite number of Slater determinants appear in the target many-body states. 
In practice, this assumption is always valid in GSM as target wave functions are always bound or resonant, so that their high-energy components in a Berggren basis are extremely small.
Hence, convergence is virtually attained in finite model spaces where all occupied one-body scattering functions bear moderate linear momentum.
As a consequence, the antisymmetry between $|n\ell_c j_c \tau_c  \rangle$ and $| \Psi_{c} \rangle^{J^\pi_{c}}$ in Eq. (\ref{coupled_Psi_rljtau}) no longer plays a role  for $n$ larger than a given $n_c^{max}$. 

Likewise, due to the finite-range property of $V_{res}-U_0$, the matrix elements $\langle \alpha \beta | V_{res}-U_0 | \gamma \delta \rangle$ vanish when $n_\alpha > n_\alpha^{max}$ (same for $\beta, \gamma$ or $\delta$). It is thus convenient to rewrite the Hamiltonian $H$ of Eq. (\ref{H_Ubasis_Vres}) introducing an operator which acts only on the target:
\begin{eqnarray}
H &=& T + U_{basis} + (V_{res}-U_0)^{A-1} + \nonumber \\
&+& [(V_{res}-U_0) - (V_{res}-U_0)^{A-1}] \ , 
\label{H_Ubasis_Vres_A_minus_one}
\end{eqnarray}
where one defines $(V_{res}-U_0)^{A-1}$ as the part of $V_{res}-U_0$ acting on the $(A-1)$-body states only for the non-antisymmetrized $A$-body states:
\begin{eqnarray}
&&\!\!\!\!\!\!\!\!\!\!(V_{res}-U_0)^{A-1} (| \Psi \rangle^{J^\pi} \otimes |n \ell j\tau \rangle) \nonumber \\
&=& [(V_{res}-U_0)  | \Psi \rangle^{J^\pi} ]\otimes |n \ell j\tau \rangle \label{Vres_restricted}\ .
\end{eqnarray}

The CC Hamiltonian kernels read using the Berggren basis expansion of projectile+target states of Eqs.   (\ref{a_dagger},\ref{channel_state_expansion}):
\begin{eqnarray}
\mathcal{H}_{cc'}^{J^\pi_A M_A} (r,r') =  \sum_{n,n'} \frac{u_{nc}(r)}{r}\frac{u_{n'c'}(r')}{r'} \; \mathcal{H}_{cc'}^{J^\pi_A M_A} (n,n')
\label{H_change_of_basis}
\end{eqnarray}
with the following definition of the hamiltonian kernels expressed in the Berggren basis
\begin{eqnarray}
\mathcal{H}_{cc'}^{J^\pi_A M_A} (n,n')  = \langle \mathcal{A} (c,n)^{J^\pi _A }_{M_A} \vert H\vert \mathcal{A}(c',n')^{J^\pi _A}_{M_A} \rangle \ . \label{H_ccp_def}
\end{eqnarray}

As antisymmetry does not play a role when one nucleon of the $A$-body system stands outside  the model space, the sum in Eq. (\ref{H_change_of_basis}) is separated into four different terms:
\begin{eqnarray}
&& \!\!\!\!\!\!\!\!\!\!\!\!\!\!\!\!\!\! \mathcal{H}_{cc'}^{J^\pi_A M_A} (r,r')   \nonumber \\
&=& \sum_{\substack{n \leqslant n_{c}^{max} \\ n' \leqslant n_{c'}^{max}}}\frac{u_{nc}(r)}{r}\frac{u_{n'c'}(r')}{r'} \; \mathcal{H}_{cc'}^{J^\pi_A M_A} (n,n')  \nonumber \\
&+& \sum_{\substack{n > n_{c}^{max} \\ n' \leqslant n_{c'}^{max}}}\frac{u_{nc}(r)}{r}\frac{u_{n'c'}(r')}{r'} \; \mathcal{H}_{cc'}^{J^\pi_A M_A} (n,n')  \nonumber \\
    &+&\sum_{\substack{n \leqslant n_{c}^{max} \\ n' > n_{c'}^{max}}}\frac{u_{nc}(r)}{r}\frac{u_{n'c'}(r')}{r'} \; \mathcal{H}_{cc'}^{J^\pi_A M_A} (n,n') 
    \nonumber \\
&+& \sum_{\substack{n > n_{c}^{max} \\ n' > n_{c'}^{max}}}\frac{u_{nc}(r)}{r}\frac{u_{n'c'}(r')}{r'} \; \mathcal{H}_{cc'}^{J^\pi_A M_A} (n,n') 
\label{H_decomposition} \ .
\end{eqnarray}

The first term in Eq. (\ref{H_decomposition}) is a finite sum and can be calculated numerically from the Slater determinant expansion of the considered many-body states using standard shell model formulas. 

The second sum can be shown to be equal to zero. This comes from the facts that all states $| n \ell_c j_c \tau_c \rangle$ with $n > n_{c}^{max}$ are orthogonal to any occupied s.p. states in target states, that $| n \ell_c j_c \tau_c \rangle$ and $| n' \ell_{c' } j_{c' } \tau_{c' } \rangle$ are orthogonal with the choice of $n$ and $n'$ in the sums, and that $H$ couples only target Slater determinants whose occupied s.p. states $| n_i \ell j \tau \rangle$ verify $n_i \leq n^{max}_{\ell j \tau}$, $i=1,.,A_{val}$. 

For the same matter, interchanging $n$ and $n'$, the third sum is equal to zero. 
Denoting $e_{nc}$ and $E_{T_c}$ respectively the s.p. energy of $| n \ell_c j_c \tau_c \rangle$ and  the energy of the target $| \Psi_{c} \rangle^{J^\pi_{c}}$, one has for $n > n_{c}^{max}$ and $n' > n_{c'}^{max}$, using Eqs. (\ref{H_Ubasis_Vres},\ref{H_ccp_def}):
\begin{eqnarray}
&& \!\!\!\!\!\!\!\!\!\! \mathcal{H}_{cc'}^{J^\pi_A M_A} (n,n') \nonumber \\
&=& \langle \mathcal{A}(c,n)^{J^\pi _A }_{M_A} \vert H \vert \mathcal{A}(c',n')^{J^\pi _A}_{M_A} \rangle \nonumber \\
&=& \langle (c,n)^{J^\pi _A }_{M_A} \vert H \vert (c',n')^{J^\pi _A}_{M_A} \rangle \nonumber \\
&=& \langle \Psi_{c}^{J^\pi_{c}} \vert H \vert \Psi_{c'}^{J^\pi_{c'}} \rangle \langle n \vert n' \rangle + \langle \Psi_{c}^{J^\pi_{c}} \vert \Psi_{c'}^{J^\pi_{c'}} \rangle \langle n \vert t + U_{basis} \vert n' \rangle   \nonumber \\
&=& \left( E_{T_c} + e_{nc} \right) ~ \delta_{cc'} ~ \delta_{nn'} \,
\end{eqnarray}
due to the disappearance of antisymmetry and the vanishing property of $V_{res}-U_0$ matrix elements when one-body states of high energy are involved.
The calculation of the last sum of Eq. (\ref{H_decomposition}) thus comes forward:
\begin{eqnarray}
&& \!\!\!\!\!\!\!\!\!\! \sum_{\substack{n > n_{c}^{max} \\ n' > n_{c'}^{max}}}\frac{u_{nc}(r)}{r}\frac{u_{n'c'}(r')}{r'} \; \mathcal{H}_{cc'}^{J^\pi_A M_A} (n,n')    \nonumber \\
&=&\delta_{cc'} \sum_{n}\frac{u_{nc}(r)}{r}\frac{u_{nc}(r')}{r'} \; \left( E_{T_c} + e_{nc}  \right) \nonumber \\
&&- \delta_{cc'} \!\!\!\!\!\sum_{n \leqslant n_{c}^{max}}\frac{u_{nc}(r)}{r}\frac{u_{nc}(r')}{r'} \; \left( E_{T_c} + e_{nc}\right) \label{2}
 \ ,
\end{eqnarray}
where a sum starting at $n = 0$ has been exposed to make completeness relations appear. Indeed:
\begin{eqnarray}
\sum_{n} \frac{u_{n}(r)}{r} \frac{ u_{n}(r')}{r'} &=& \frac{\delta(r- r')}{rr'}  \nonumber \\
\sum_{n} \frac{u_{n}(r)}{r}e_{cn}\frac{ u_{n}(r')}{r'}&=& [T(r) + U_{basis}(r)]\,\frac{\delta(r- r')}{rr'} \nonumber
\end{eqnarray}
and the CC Hamiltonian kernels become:
\begin{eqnarray}
&& \!\!\!\!\!\!\!\!\!\! \mathcal{H}_{cc'}^{J_A M_A} (r,r') \nonumber \\
& = & \delta_{cc'} \left[ -\frac{\hbar ^2}{2\mu_c} \frac{1}{r}\frac{\partial ^2}{\partial r^2} r + \frac{\hbar ^2 \ell_c (\ell_c +1)}{2\mu_cr^2} +  E_{T_c} \right] \frac{\delta(r-r')}{rr'}  \nonumber    \\
&+&    \delta_{cc'} \frac{\delta(r-r')}{rr'}  U_{basis}(r) + \tilde{V}_{cc'}^{J_A M_A}(r,r')\ .
\end{eqnarray}
$\tilde{V}_{cc'}^{J_A M_A}$ stands for the remaining short-range potential terms of the Hamiltonian kernels, i.e. the first sum of Eq. (\ref{H_decomposition})  and the last sum of Eq. (\ref{2}).

\bigskip

The derivation of the overlap kernels Eq. (\ref{overlap_kernels}) is similar to that of the Hamiltonian kernels, replacing the Hamiltonian operator by the identity and leads to:
\begin{eqnarray}
\mathcal{O}_{cc'}^{J_A M_A} (r,r') & = & \delta_{cc'} \frac{\delta(r-r')}{rr'}  + \tilde{O}^{J_A M_A}_{cc'}(r,r')\ .
\end{eqnarray}
The overlap kernels are thus the sum of the identity and a short-range exchange term $\tilde{O}_{cc'}^{J_A M_A}$ coming from the non-orthogonality of the channels. It is important to note that it is the short-range property of the non-orthogonality of the channels that justifies the uniqueness of the expansion (Eq. (\ref{10})) of the scattering state.

\bigskip

 The use of harmonic oscillator representation is numerically advantageous as compared to coordinate representation for the short-range part of the hamiltonian and overlap kernels. For that, due to completeness properties of both coordinate and harmonic oscillator basis expansions, it is sufficient to replace 
 $$\frac{u_{nc}(r)}{r} = \langle n\ell_c j_c \tau_c | r \ell_{c} j_{c} \tau_c\rangle $$ by the overlap $\langle n\ell_c j_c \tau_c | \alpha \ell_{c} j_{c} \tau_c\rangle$, where $|\alpha \ell_{c} j_{c} \tau_c\rangle$ is a harmonic oscillator basis state, and sum over $\alpha$   in the  relations.

\section{Resolution of the CC equations\label{sec3}}

So far, the problem as expressed in Eq. (\ref{cc_eqs_formal_radial}) is a generalized eigenvalue problem which can take the matrix form: 
\begin{equation} 
\mathcal{H} U = E \mathcal{O} U .
\end{equation}
 $U = \{ u_c(r) \}_c$ is the vector of the radial amplitudes. Introducing $W = \mathcal{O}^{\frac{1}{2}} U$, and the modified Hamiltonian: 
\begin{equation}
\mathcal{H}_m = \mathcal{O}^{-\frac{1}{2}} \mathcal{H} \mathcal{O}^{-\frac{1}{2}} \ ,
\label{Hm}
\end{equation}
 one obtains the standard eigenvalue problem: \begin{equation}
 \mathcal{H}_m W = E W \ ,
 \end{equation}
  where $W = \{ w_c(r) \}_c$ is now to be determined.

Let us define $\Delta $ as the finite-range part of $\mathcal{O}^{-\frac{1}{2}}$, i.e. $$\Delta = \mathcal{O}^{-\frac{1}{2}} - \Id \ .$$
It follows that $\mathcal{H}_m$ can be separated into long- and short-range parts: 
\begin{equation}
\mathcal{H}_m = \mathcal{H} + \mathcal{H} \Delta + \Delta \mathcal{H} + \Delta \mathcal{H} \Delta \ ,
\label{Hm1}
\end{equation}
where all terms involving $\Delta$ are expanded in the harmonic oscillator basis. Thus, the added part of $\mathcal{H}$ in $\mathcal{H}_m$ can be treated similarly to the short-range residual interaction.

Using results of Sec. \ref{cc_eqs_derivation} and the transformation described above, Eq. (\ref{cc_eqs_formal_radial}) can be written as a system of non-local differential equation with respectively local $V^{(loc)}_{c}(r)$ and non-local $V^{(non-loc)}_{c c'}(r,r')$ optical potentials:
\begin{eqnarray}
&& \!\!\left[ -\frac{\hbar ^2}{2\mu_c} \frac{d \:^2}{d r^2}  + \frac{\hbar ^2\ell_c (\ell_c +1)}{2\mu_cr^2} \: + V^{(loc)}_{c}(r) \right] w_{c} (r) \nonumber \\
&  & +  \sum_{c'} \int_0^{+\infty} \!\!\!V^{(non-loc)}_{c c'}(r,r') \: w_{c'}(r') \: d r' \nonumber\\
&  &  \quad\quad = \left( E - E_{T_c}\right) w_{c} (r) \ .
\label{cc_eqs_diff_radial}
\end{eqnarray} 
Once the solution $W = \{ w_c(r) \}_c$ of Eq. (\ref{cc_eqs_diff_radial}) is determined, the initial vector $U = \{ u_c(r) \}_c$ of channel functions is given by $U =  \mathcal{O}^{-\frac{1}{2}} W $ or
$$U = W + \Delta  W\ ,$$ which reads using coupled-channel representation:
\begin{equation}
u_c(r) = w_c(r) + \sum_{c'}  \int_0^{+\infty} \!\!\!\Delta _{cc'}(r,r') w_{c'}(r') \: d r' \label{uc_from_wc}.
\end{equation}
Normalization of the full coupled-channel state $| \Phi \rangle$ remains to be effected. For this, if $| \Phi \rangle$ is a bound or resonant state, one calculates the squared norm of $| \Phi \rangle$:
\begin{align}
\langle \Phi | \Phi \rangle &= \sum_{cc'} \langle u_c | \mathcal{O}_{cc'} | u_{c'} \rangle \nonumber \\
                            &= \sum_{c \!\!} \int u_c^2(r)~ dr + \sum_{cc'} \langle u_c | \mathcal{O}_{cc'} \!-\! \Id_{cc'} | u_{c'} \rangle \label{Phi_norm}.
\end{align}
The first term of Eq. (\ref{Phi_norm}) is calculated using complex scaling and the second term using the harmonic oscillator expansion of the $\mathcal{O}- \Id$ operator, as it is finite-range.

In the case of scattering states, one simply demands that the incoming part of $W(r)$ in the entrance channel $c_0$ is normalized
to unity, i.e. $$w^{(-)}_{c_0}(r) = H^-_{\ell_{c_0}}(\eta_{c_0},k_{c_0} r)$$ in the asymptotic region, with $\ell_{c_0}$, $\eta_{c_0}$ and $k_{c_0}$ 
the orbital angular momentum, Sommerfeld parameter and linear momentum of the entrance channel $c_0$, respectively.

\subsection{Equivalent potential method for solving GSM integro-differential equations} \label{eq_pot_GSM}

In order to deal with a local problem, Eq. (\ref{cc_eqs_diff_radial}) is rewritten as:
\begin{eqnarray}
w_c''(r) &=& \left( \frac{\ell_c(\ell_c+1)}{r^2} - k_c^2 \right) w_c(r) \nonumber \\
&+&\frac{2 \mu_c}{\hbar^2} \sum_{c'} V^{(eq)}_{cc'}(r) w_{c'}(r) + \frac{2 \mu_c}{\hbar^2} S_c(r)
\label{cc_local_eqs}
\end{eqnarray}
where $k_c^2 = (2 \mu_c/\hbar ^2)(E - E_{T_c})$, $V^{(eq)}_{cc'}(r)$ is the equivalent local potential and $S_c(r)$ is an additional source term, both depending on the channel wave functions and defined by:
\begin{align}
V^{(eq)}_{cc'}(r) &= V^{(loc)}_{c}(r)\cdot\delta_{cc'}   \nonumber \\
&\;+  \frac{1 - F_{c'}(r)}{w_{c'}(r)}\!\!\int V^{(non-loc)}_{cc'}(r,r') w_{c'}(r')\:dr' \! \label{Veq}  \\ 
S_c(r) &= \sum_{c'} F_{c'}(r)\!\!\int V^{(non-loc)}_{cc'}(r,r') w_{c'}(r')\:dr'  .  \label{source}
\end{align}
$F_c(r)$ is a smoothing function, which we will detail afterwards.
Eq. (\ref{cc_local_eqs}) is the generalization of the one-dimensional equivalent potential method described in Ref. \cite{Mic09x}.
As in the one-dimensional case, the naive equivalent potential method would consist in posing $F_c(r) = 0$ in Eqs.   (\ref{Veq},\ref{source}). However, this would imply the divergence of $V^{(eq)}_{cc'}(r)$ when $w_{c'}(r) = 0$,
so that the equivalent local potentials become singular. In order to avoid this situation, one utilizes a smoothing function which cancels out divergences at the zeroes of $w_{c'}(r)$, and one introduces a source term $S_c(r)$ in Eq. (\ref{cc_local_eqs}) so that the local problem remains
equivalent to the non-local problem of Eq. (\ref{cc_eqs_diff_radial}). One demands that $F_c(r)$ cancels out the divergences of $1/w_c(r)$, i.e. close to the zeroes of $w_c(r)$, except at $r=0$, as one can show that no singularity can occur at this point; $F_c(r) \sim 0$ elsewhere. The ansatz: 
\begin{eqnarray}
F_c(r) &=& \exp \left( -\alpha \left| \frac{w_c(r)}{w'_c(r)} \right|^2 \right) \cdot \nonumber \\
&&\cdot \left( 1 - \exp \left[ -\alpha \left| \frac{w_c^{(asymp)}(r)}{w_c(r)} - 1 \right|^2 \right] \right) \label{Fc}
\end{eqnarray}
is used, where $\alpha$ is typically chosen between 10 and 100, and $w_c^{(asymp)}(r)$ is the asymptotic form of $w_c(r)$ for $r \sim 0$, which becomes rapidly different from $w_c(r)$ when $r$ increases, so that Eq. (\ref{Fc}) fulfils the above requirements.

The use of the $w_c(r)/w'_c(r)$ ratio in Eq. (\ref{Fc}) ensures that $F_c(r) \sim 0$ when $r \rightarrow +\infty$, as one wants $F_c(r) \sim 1$ only close to the finite zeroes of $w_c(r)$.
$V^{(eq)}_{cc'}(r)$ and $S_c(r)$ are also multiplied by cut functions for $r \rightarrow +\infty$, as they often decrease too slowly with $r$, 
but we do not consider it in Eqs. (\ref{Veq},\ref{source}) for simplicity.

As $V^{(eq)}_{cc'}(r)$ and $S_c(r)$ depend on the $w_c(r)$ channel wave functions which one calculates, they have to be determined iteratively, as in a Hartree-Fock procedure.
For this, starting values for $V^{(eq)}_{cc'}(r)$ and $S_c(r)$ are chosen (their determination will be delineated afterwards), Eq.~(\ref{cc_local_eqs}) is solved, the new  $V^{(eq)}_{cc'}(r)$ and $S_c(r)$
functions are calculated from the obtained $w_c(r)$ channel wave functions, and the process is continued until convergence.

However, during the iterative process, Eq. (\ref{cc_local_eqs})
does not form a symmetric problem, that is $$V^{(eq)}_{cc'}(r) \neq V^{(eq)}_{c'c}(r) \, .$$ While this is not relevant when one calculates scattering states, whose linear momenta $k_c$ are fixed,
it becomes critical when one calculates bound or resonant coupled-channel wave functions, as then bound states no longer have real energies and resonant states positive widths ensuing in the divergence of the iterative process. It is thus necessary to symmetrize Eq. (\ref{cc_local_eqs}) in this case. One uses the symmetrized equivalent potentials
and sources $V^{(eq,sym)}_{cc'}(r)$ and $S_c^{(sym)}(r)$ defined by:
\begin{align}
V^{(eq,sym)}_{cc'}(r) &= \frac{V^{(eq)}_{cc'}(r) + V^{(eq)}_{c'c}(r)}{2} \mbox{ for } c' \neq c \label{Veq,sym_off_diag} \\
V^{(eq,sym)}_{cc}(r) &= V^{(eq)}_{cc}(r) \nonumber \\
& \!\!\!\!\!\!\!\!\!\!\!+ \frac{1 - F_c(r)}{w_c(r)} \!\sum_{c' \neq c} \frac{V^{(eq)}_{cc'}(r) - V^{(eq)}_{c'c}(r)}{2} w_{c'}(r)  \label{Veq,sym_diag} \\
S_c^{(sym)}(r) &= S_c(r)  \nonumber \\
+& \: F_c(r) \sum_{c' \neq c} \frac{V^{(eq)}_{cc'}(r) - V^{(eq)}_{c'c}(r)}{2} w_{c'}(r) \label{source-sym}
\end{align}
where one can verify that the newly defined values $V^{(eq,sym)}_{cc'}(r)$ and $S_c^{(sym)}(r)$ also render the problem equivalent to the initial integro-differential problem when inserted in Eq. (\ref{cc_local_eqs}).

As the method is embedded in an iterative procedure, it is necessary to generate a good starting point for it to converge. For that purpose, Eq. (\ref{cc_eqs_diff_radial}) is firstly diagonalized with the Berggren basis used in the Gamow Shell Model calculation of targets eigenstates. 

If one aims at bound or resonant coupled-channel wave functions,
the bound or resonant states obtained from the Berggren basis diagonalization provide the starting $w_c^{(Berggren)}(r)$ coupled-channel wave functions, which in their turn determine the
starting $V^{(eq)}_{cc'}(r)$ and $S_c(r)$ functions. 

In the case of scattering coupled-channel wave functions, one chooses from the Berggren basis diagonalization the eigenstate $w_c^{(Berggren)}(r)$
whose energy is closest to the one which one considers in Eq. (\ref{cc_eqs_diff_radial}), so that starting $V^{(eq)}_{cc'}(r)$ and $S_c(r)$ functions can be implemented as well.
It has been noticed in practice that these starting $V^{(eq)}_{cc'}(r)$ and $S_c(r)$ functions are very close to the exact ones, so that the iterative procedure converges rapidly
using the aforementioned starting point.

\subsection{Boundary conditions and basis functions} \label{basis_functions}

Precise boundary conditions of the wave functions have to be specified for the numerical integration. In the vicinity of zero, the source terms in the coupled-channel equations are equal to zero and Eq. (\ref{cc_local_eqs}) reads:
\begin{equation}
w_c''(r) \; \sim \left( \frac{\ell_{c}(\ell_{c}+1)}{r^2} + a_c \right)\!w_c(r) +  \sum_{c' \neq c} a_{cc'} w_{c'}(r)  \label{cc_eqs_zero_gen}
\end{equation}
where $$a_c = (2\mu_c/\hbar^2) V^{(eq)}_{c c}(0) - k_c^2$$ and $$a_{cc'} = (2\mu_c/\hbar^2) V^{(eq)}_{c c'}(0)~ \ .$$ 

Due to the presence of the coupling terms, one does not always have the standard behavior $w_c (r) \sim r^{\ell_c +1}$. As these behaviors are undetermined, we utilize forward basis functions $W_b^{(0)}(r) = \{ w^{(0)}_{bc}(r) \}_c$ indexed by the letter $b$, which exhibit simplistic behaviors, that is $w^{(0)}_{bc}(r) \sim C_{b}^{(0)}  r^{\ell_b + 1} $ for $c=b$ and 
$w^{(0)}_{bc}(r) = o (r^{\ell_b + 1} )$ otherwise. $C_{b}^{(0)}$ denotes a constant that is determined in the previous iteration in the iterative process. It is immediate to verify that for $c\neq b$, the behavior of $w^{(0)}_{bc}(r) = o (r^{\ell_b + 1} )$ can be separated as:
\begin{equation}
w^{(0)}_{bc}(r) \sim \:  \frac{a_{cb}}{(\ell_{b} + 2)(\ell_{b} + 3) - \ell_{c}(\ell_{c} + 1)}  C_{b}^{(0)}r^{\ell_{b} + 3}   \\
\end{equation}
for $\ell_c\neq \ell_b +2$ and
\begin{equation}
w^{(0)}_{bc}(r) \sim \: \frac{a_{cb}}{2\ell_{b} + 5} C_{b}^{(0)} r^{\ell_{b} + 3} \log(r) 
\end{equation}
for $\ell_{c} = \ell_{b} + 2$. The basis functions $W_b^{(0)}(r)$ are thus numerically integrated starting from $r=0$ with these boundary conditions. However, due to the inhomogeneous character of the CC equations, a modified version of Eq. (\ref{cc_local_eqs}) has to be used for the integration. Indeed, the full solution  $W(r)$ will be searched among the linear combinations:
\begin{eqnarray}
W(r) & = & A_b^{(0)} W_b^{(0)}(r) \label{forward_expansion}  \ ,
\end{eqnarray}
implying that the source term $S_c(r)$ must be handled in a special manner. For bound and resonant states, one divides the source term $S_c(r)$ by the number of channels in Eq. (\ref{cc_local_eqs}). For scattering states, the source term $S_c(r)$ is suppressed in the calculation of $W_b^{(0)}(r)$ and will be considered only in the incoming part of the wave functions as explained afterwards. 

At large distances, the boundary conditions are straightforward, but will be expressed  by means of the backward basis functions in a similar manner to what is done in the vicinity of zero essentially for numerical reasons. Let us define the backward basis functions $$W_b^{(+)}(r) = \{ w^{(+)}_{bc}(r) \}_c$$ which verify $w^{(+)}_{bc}(r) \sim  C_{b}^{(+)} H^+_{\ell_{b}}(\eta_{b},k_{b} r)$ if $c=b$ and $w^{(+)}_{bc}(r) \sim 0$ otherwise, $C_{b}^{(+)}$ being a constant determined during the previous iteration. The vector of radial amplitudes $W(r)$ is looked upon the combinations:
\begin{eqnarray}
W(r) & = & A_b^{(+)} W_b^{(+)}(r) + W^{(-)}(r)\label{backward_expansion}\ ,
\end{eqnarray}
where $W^{(-)}(r)$ stands for the incoming part of $W(r)$, which is identically equal to zero if the state is bound or resonant. In the case of a scattering state, it verifies $w^{(-)}_{c_0}(r) \sim H^-_{\ell_{c_0}}(\eta_{c_0},k_{c_0} r)$ in the incoming channel $c_0$ and $w^{(-)}_{c}(r) \sim 0$ for $c \neq c_0$   in the asymptotic region.  $W_b^{(+)}(r)$ and $W^{(-)}(r)$ are integrated from a starting point $r=R_{max}$ to the decreasing radii, in which $R_{max}$ is chosen to be a radius after which the nuclear interaction vanishes (one typically selects $R_{max}=15$ fm). 

As pointed out in the forward basis case, $W_b^{(+)}(r)$ and $W^{(-)}(r)$ cannot  be solutions of Eq. (\ref{cc_local_eqs}), as its non-homogeneous character implies that $W(r)$ as defined in Eqs.  (\ref{backward_expansion}) would not be another solution. For the bound and resonant cases, the source term $S_c(r)$ in Eq. (\ref{cc_local_eqs}) is thus divided by the number of channels. In the case of scattering states, the source term $S_c(r)$ is put to zero in the definition of $W_b^{(+)}(r)$ and each $w^{(-)}_{c}(r)$ for $c\neq c_0$, and is only taken into account when calculating $w^{(-)}_{c_0}(r)$. Using this modification of the source term $S_c(r)$ for the computation of $W_b^{(0)}(r)$, $W_b^{(+)}(r)$ and $W^{(-)}(r)$, one can check that the full coupled-channel wave function $W(r) = \{ w_c(r) \}_c$ verifies Eq. (\ref{cc_local_eqs}).

 \subsection{Solution of the coupled-channel equations\label{sec3c}}

The matching of the linear combinations of the two sets of basis wave functions at a given radius $R$ will provide the solution of the coupled-channel equations. 
As all channel wave functions $w_c(r)$ must be continuous and have their derivatives continuous, 
one obtains the following equations using Eqs. (\ref{forward_expansion},\ref{backward_expansion}):
\begin{equation}
\begin{array}{rl}
\displaystyle \sum_b \!\!\left[A_b^{(0)} w_{bc}^{(0)}(R)  - A_b^{(+)} w_{bc}^{(+)}(R) \right] \!\!\!&\displaystyle = w_{c}^{(-)}(R)   \\
\vspace*{-0.1in}& \\
\displaystyle\sum_b \!\!\left[A_b^{(0)} \frac{dw_{bc}^{(0)}}{dr}\!(R)  - A_b^{(+)} \frac{dw_{bc}^{(+)}}{dr}\!(R) \right]  \!\!\!& \displaystyle =\!\frac{dw_{c}^{(-)}}{dr}\!(R)\, .
\end{array} 
\!\!
\label{c0_match_der_eqs}
\end{equation}

For scattering states, Eq. (\ref{c0_match_der_eqs}) form a linear system $AX = B$, immediate to solve. 
In practice, $R \sim  2 - 5$~fm, that is close to the surface of the nucleus, 
so that $w_{bc}^{(0)}(r)$ and $w_{bc}^{(\pm)}(r)$ functions are integrated respectively forward and backward from their asymptotic region up to a radius where they are close to one in modulus.

For bound and resonant states,  Eq. (\ref{c0_match_der_eqs}) form a $AX=0$ system, as there is no incoming channel. 
Hence, in this case, one has to solve $\det A = 0$, which occurs only if $W(r)$ has a bound or resonant eigenenergy.
$\det A$ is thus the generalization of the Jost function for coupled-channel equations. Once the eigenenergy for which $\det A = 0$ has been found, the constants $A_b^{(0)},A_b^{(+)}$ are given by the eigenvector $X$ of $A$ of zero eigenvalue.

Due to the presence of $C_{b}^{(0)}$ and $C_{b}^{(+)}$ in the asymptotic functions, the convergence of the iterative process is attained when $A_b^{(0)}$ and $A_b^{(+)}$ are sufficiently close to one, yielding the orthogonalized radial amplitude wave functions $w_{c}^{J^\pi _A}(r)$ and subsequently the initial radial amplitudes $u_{c}^{J^\pi _A}(r)$ using Eq. (\ref{uc_from_wc}). The calculations of scattering states for different entrance channels $c_0$ and spin-parity $J^\pi_A$ allow the extraction of the $S$-matrix elements $S^{J^{\pi}_A}_{c_0c}$  from the asymptotic constants $C_c^{(+)}$ given in Sec. \ref{basis_functions}. The determination of scattering-related physical observables naturally comes forward.

\section{GSM-CC calculation of reaction cross-sections\label{results}}

In the calculation of $^{18}$Ne$(p,p')$ elastic and inelastic cross-sections, the reaction channels are defined by the ground state $J^{\pi}=0_1^+$ and the first excited state $J^{\pi}=2_1^+$ of $^{18}$Ne target coupled to the proton in different partial waves  with $\ell_c\leq 2$. 
The s.p. proton configuration corresponds to $0d_{5/2}$ and $1s_{1/2}$ resonances and 28 states of a discretized contour for each resonance. Moreover, we include partial waves $(d_{3/2}, p_{3/2}, p_{1/2})$ which are decomposed using a real-energy contour of 21 points as in this case the resonant poles at high energy are very broad.

\subsection{The potential of $^{16}$O core\label{potential}}

The Hamiltonian consists of the Woods-Saxon potential describing the $^{16}$O core and the effective two-body interaction among valence nucleons. Parameters of the core potential: the radius $R_0=3.05$ fm, the depth of the central part $V_0=55.5238$ MeV, the diffuseness $d=0.65$ fm, and the spin-orbit strength $V_{so}=6.149$ MeV have been chosen to reproduce $^{17}$O and $^{17}$F spectra. The radius of the Coulomb potential in this calculation is $R_C=3.05$~fm.

\subsection{The effective two-body interaction\label{interaction}}

The s.p. basis in $^{18}$Ne and $^{19}$Na is generated by the same Gamow-Hartree-Fock potential. This potential is obtained from the Woods-Saxon potential of the core and the effective two-body interaction between valence nucleons. In the calculations, we use the MSG\cite{GSM1} effective two-body interaction: 
\begin{eqnarray}
&&\!\!\!\!\!\!\!\!\! V_{JT}^{({\rm MSG})} (\vec{r}_1,\vec{r}_2)  =  V_0(J,T) \exp \left[ - \left( \frac{r_1-R_0}{\mu_I}\right) ^2\right]\cdot \nonumber \\
&\cdot&\exp \left[ - \left( \frac{r_2-R_0}{\mu_I}\right) ^2\right]\cdot \nonumber \\
&\cdot& F(R_0,r_1) F(R_0,r_2) \sum_{\ell m}  {Y^\ell_m}^{*} (\Omega_1) Y^\ell_m (\Omega_2) ,
\label{AutresModeles::MSGI}
\end{eqnarray}
where $F(R_0,r)$ is a Fermi function:
\begin{equation}
	 F(R_0,r) \; = \; \frac{1}{1+\exp\left[(r-2R_0+r_F) / d_F\right]} \;\:.
	 \label{Autres_Modeles::Fermi}
\end{equation}
In Eq. (\ref{Autres_Modeles::Fermi}), $r_F =  1$ fm and the diffuseness parameter is $d_F=0.05~{\rm fm}\ll R_0$, so that the MSG interaction is centered at the nuclear surface and becomes negligible for $r \gtrsim 2R_0$.
\begin{table} [htb]
\caption{\label{tab:1}Parameters (in units MeV$\cdot$fm$^3$) of the MSG interaction which have been adjusted to reproduce spectra of $^{18}$Ne and $^{19}$Na. }
\begin{ruledtabular}
\begin{tabular}{|c||c|c|c|c|c|c|}
$J$ & 0 &1 & 2 & 3 & 4 & 5  \\ \hline
$V_0(J,T=1)$ & -7.42 & 0 & -4.65 & 8.65 & 0.74 & 0 \\
\end{tabular}
\end{ruledtabular}
\end{table}
Parameters of the MSG interaction (see Table \ref{tab:1}) have been fitted to reproduce in GSM the excitation energies of low-lying states of $^{18}$Ne and $^{19}$Na with respect to the $^{18}$Ne ground state. 
\begin{table} [htb]
\caption{\label{tab:2} Comparison between the experimental data and the GSM results for the two-proton separation energy $S_{2p}$, and the low-energy states of $^{18}$Ne. GSM calculation has been done using the MSG two-body interaction with the parameters given in Table \ref{tab:1}.  }
\begin{ruledtabular}
\begin{tabular}{|c|c|c|}
 $^{18}$Ne  & EXP & GSM    \\ \hline 
& &   \\
S$_{2p}$ [MeV] & -4.522 & -4.232    \\
& &   \\ \hline
& &   \\
$E(0_1^+)$ [MeV] & 0.000 & 0.000    \\
$E(2_1^+)$ [MeV]& 1.887 & 1.900    \\
& &   \\
\end{tabular}
\end{ruledtabular}
\end{table}
\begin{table} [htb]
\caption{\label{tab:3} Comparison between the experimental data and the GSM and GSM-CC results for the low-energy states of $^{19}$Na using the MSG two-body interaction with the parameters given in Table \ref{tab:1}.}
\begin{ruledtabular}
\begin{tabular}{|c|c|c|c|}
 $^{19}$Na  & EXP & GSM  & GSM-CC  \\ \hline 
& & &  \\
$E(5/2_1^+)$ [MeV]& 0.321 & 0.338  &  0.347 \\
$E(3/2_1^+)$ [MeV]& 0.441 & 0.448  &  0.478 \\
$E(1/2_1^+)$ [MeV]& 1.067-i0.05$\pm$0.01 & 1.065-i0.067  & 1.115-i0.079 \\
& & &  \\
\end{tabular}
\end{ruledtabular}
\end{table}

In Tables \ref{tab:2} and \ref{tab:3} we compare the GSM excitation energies of low-lying states of $^{18}$Ne and $^{19}$Na with their experimental values. For this interaction, the two-proton separation energy $S_{2p}$ in $^{18}$Ne differs from the experimental value by about 300 keV.  All states in $^{19}$Na are narrow resonances which decay by proton emission. The last column in Table \ref{tab:3} contains excitation energies of $^{19}$Na calculated in GSM-CC using the the channel states generated by the ground state $J^{\pi}=0_1^+$ and the first excited state $J^{\pi}=2_1^+$ states of $^{18}$Ne coupled to the proton in partial waves  with $\ell_c\leq 2$. One may notice a small difference between GSM and GSM-CC excitation energies of $^{19}$Na resonances.  This difference indicates the tiny lack of the many-body completeness in GSM-CC calculations, due to neglecting higher-lying discrete and scattering states of $^{18}$Ne above the first excited state $J^{\pi}=2_1^+$ in constructing the reaction channels (\ref{channel_states}).

\subsection{$^{18}$Ne(p,p') reaction cross-sections\label{detres}}

The GSM-CC equations  (\ref{cc_eqs_diff_radial}) are solved using the multidimensional iterative procedure (Sec. \ref{eq_pot_GSM}) where the input functions for the first iteration come from the diagonalization of the modified Hamiltionian $\mathcal{H}_m$ (Eq. (\ref{Hm})) in the Berggren basis. Therefore, the discretization density of the Berggren basis becomes an essential ingredient of the GSM-CC calculations assuring good convergence properties of the calculated resonances and scattering wave functions. Below, we shall present tests of the numerical method of finding the radial solution of the GSM-CC equations.

\subsubsection{The test of the method of solving the CC equations}

\begin{figure}[hbt]
\center
\includegraphics[width=0.5\textwidth]{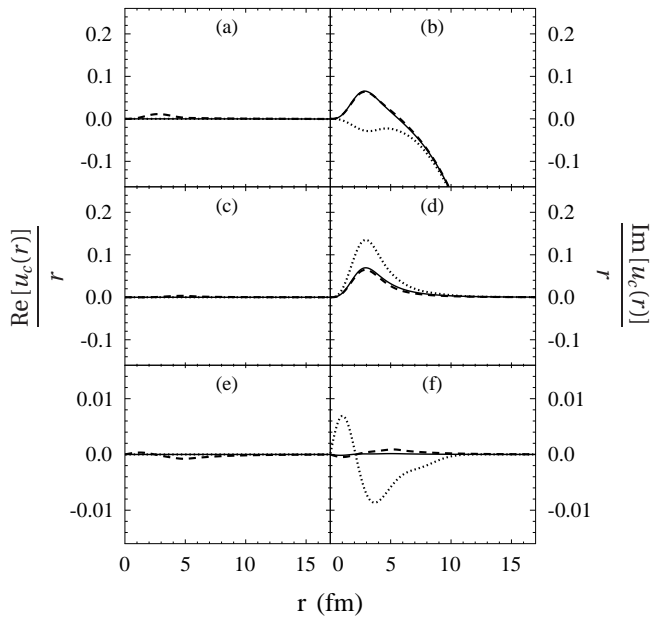}
\caption{\label{fig1} Real (left) and imaginary (right) parts of the channel wave functions for $J^\pi_A=5/2^+$ at $E_{\rm CM}=$ 1 MeV. Dashed lines show the channel functions at the first iteration. The final solution for channel wave functions is shown with the solid lines. The dotted lines depict the final solutions for the rescaled channel wave function $u_c$ in Eq. (\ref{uc_from_wc}). Panels (a) and (b) represent the entrance channel: $(0^+ \times d_{5/2})$. Panels (c), (d) correspond to the channel $(2^+ \times d_{5/2})$, and panels (e), (f) to the channel $(2^+ \times s_{1/2})$.}
\end{figure}

\begin{figure}[hbt]
\center
\includegraphics[width=0.5\textwidth]{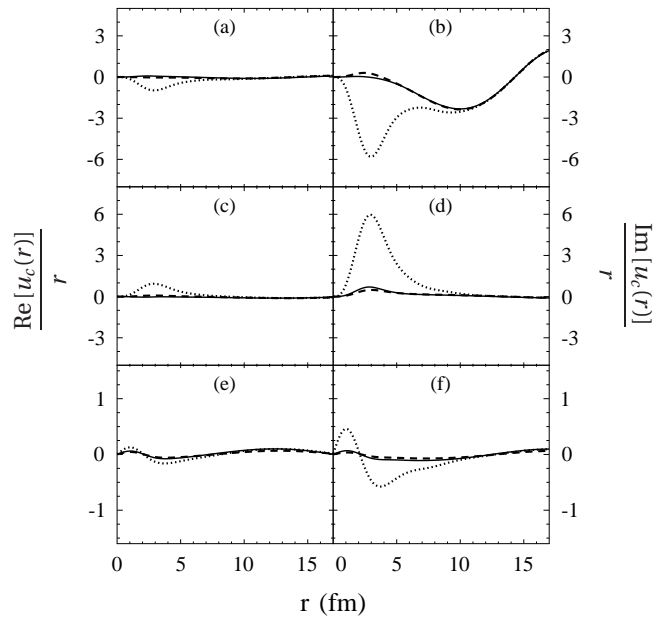}
\caption{\label{fig2} The same as in Fig. \ref{fig1} but for  $E_{\rm CM}=$ 5 MeV. }
\end{figure}

\begin{figure}[hbt]
\center
\includegraphics[width=0.5\textwidth]{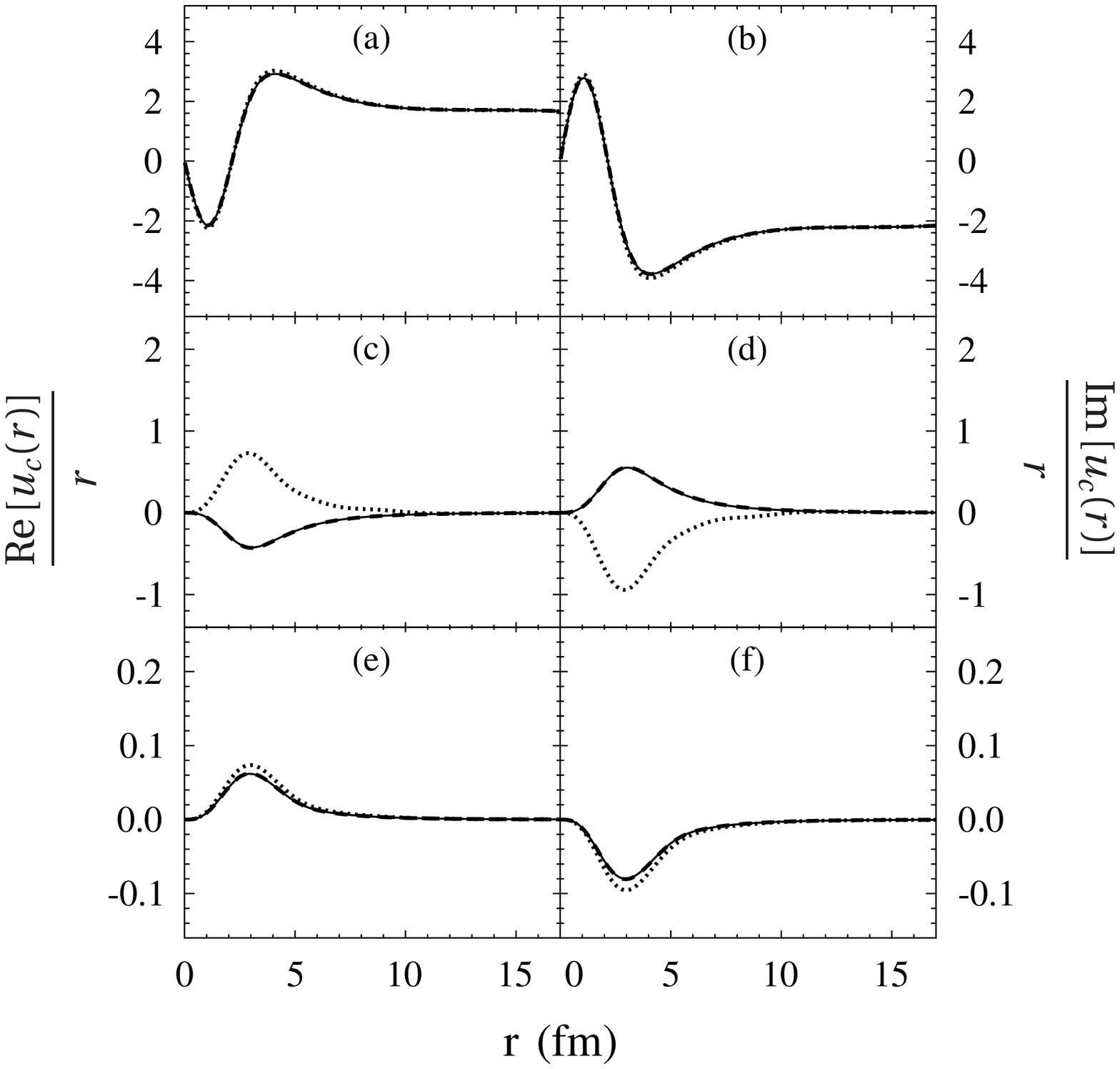}
\caption{\label{fig3} The channel wave functions for $J^\pi_A=1/2^+$ at $E_{\rm CM}=$~1~MeV.  Panels (a) and (b) show the entrance channel $(0^+ \times s_{1/2)}$. Panels (c), (d) correspond to the channel $(2^+ \times d_{5/2})$, and panels (e), (f) to the channel $(2^+ \times s_{1/2})$. For further information, see the caption of Fig. \ref{fig1}.}
\end{figure}

\begin{figure}[hbt]
\center
\includegraphics[width=0.5\textwidth]{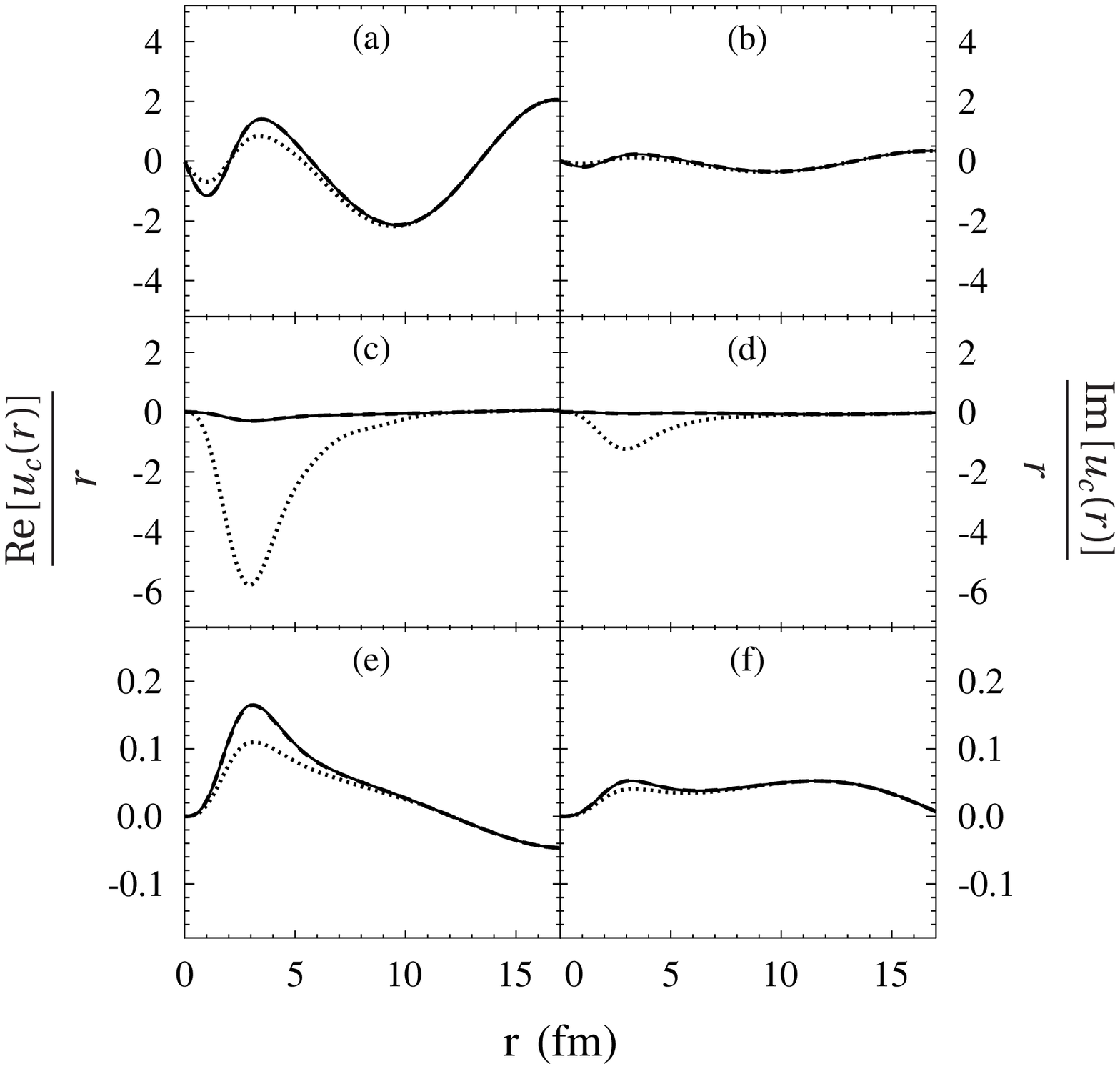}
\caption{\label{fig4} The same as in Fig. \ref{fig3} but for $E_{\rm CM}=$ 5 MeV.}
\end{figure}

Figs.~\ref{fig1}-\ref{fig4} compare the radial dependence of channel functions for different CM energies and angular momenta $J^{\pi}_A$ in the system $p$+$^{18}$Ne. 
In each considered case, the three channel functions are plotted to demonstrate the variation from an initial condition to the final solution in the iterative procedure of solving the CC equations. For each of these three cases, we present (i) the channel functions at the first iteration (the dashed lines), obtained by diagonalizing the Hamiltonian $\mathcal{H}_m$, (ii) the final solution for channel wave functions after solving the CC equations (\ref{cc_eqs_diff_radial}) (the solid lines), and (iii) the the final rescaled wave functions $u_c(r)$ (see Eq. (\ref{uc_from_wc})) (the dotted line). Both real and imaginary parts of the respective functions are shown.

Fig. \ref{fig1} is obtained for $J^{\pi}_A=5/2^+$ and $E_{\rm CM}=$ 1 MeV. The channel wave functions correspond to the entrance channel $(0^+ \times d_{5/2})$ (panels (a), (b)), the channel $(2^+ \times d_{5/2})$ (panels (c), (d)), and the channel $(2^+ \times s_{1/2})$ (panels (e), (f)). Both real (panels (a), (b), (c)) and imaginary (panels (d), (e), (f)) parts of the channel functions are depicted in the figure. One can see that in all considered cases, the channel function at the first iteration resembles closely the converged solution of the CC equations.

Fig. \ref{fig2} shows the channel functions for $J^{\pi}_A=5/2^+$ and $E_{\rm CM}=$ 5 MeV. Again, the channel functions at the first iteration and after solving the CC equations are quite similar. We may notice that the difference between the channel function $w_c$ and the rescaled channel function $u_c$ increases when increasing the total energy $E_{\rm CM}$. 

Fig. \ref{fig3} and \ref{fig4} picture the channel functions for $J^{\pi}_A=1/2^+$ at $E_{\rm CM}=$ 1 MeV and 5 MeV, respectively. Also in this case, the channel wave function at the first iteration is very close to the final solution of the CC equation. 

\begin{figure}[hbt]
\center
\includegraphics[width=0.5\textwidth]{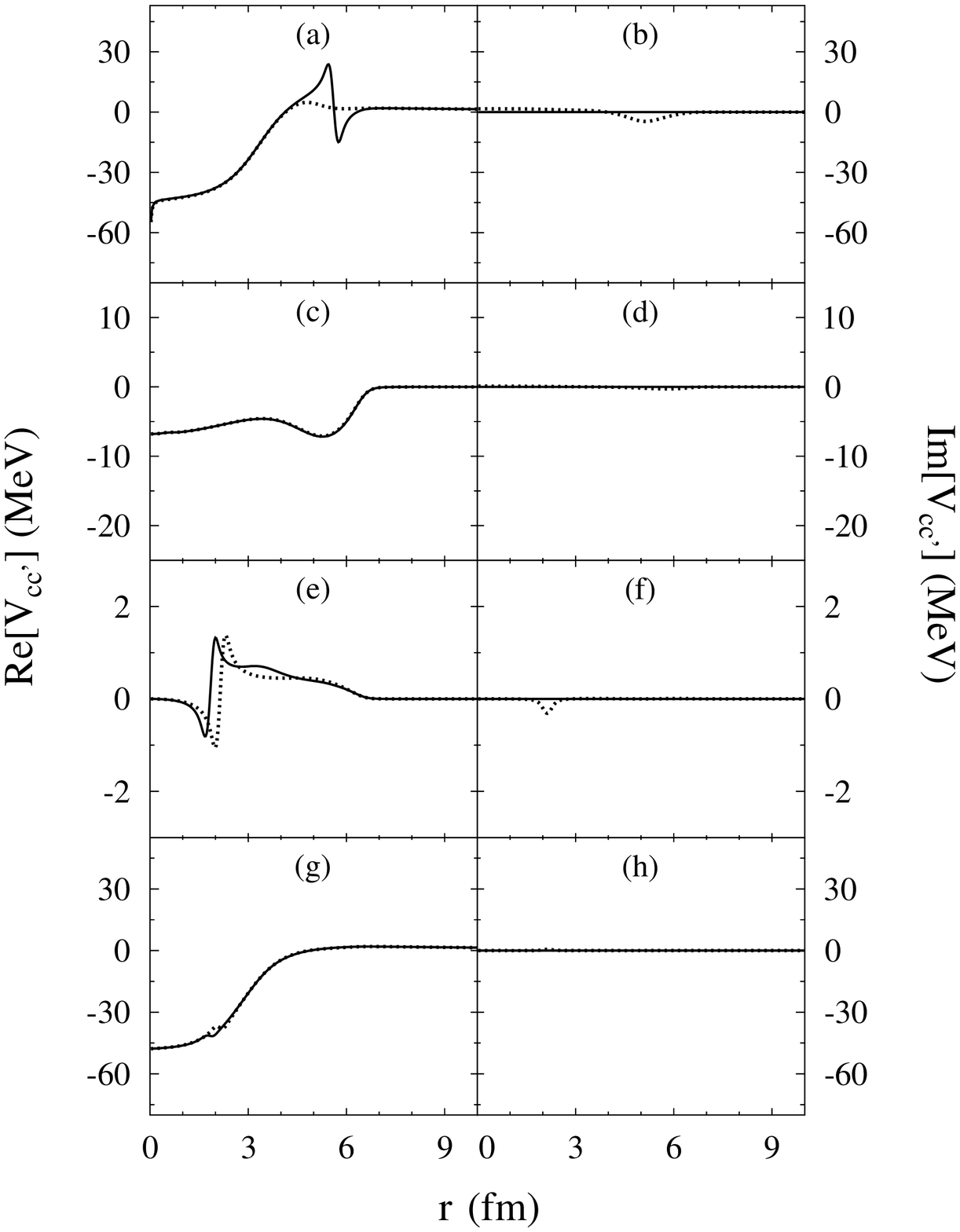}
\caption{\label{fig5} The equivalent local potentials $V_{cc'}^{(eq)}$ (Eq. (\ref{eq_pot_GSM})) for $J^\pi_A=5/2^+$ at $E_{\rm CM}=1$ MeV. The channel indices 0, 1, 2 denote the entrance channel $(0^+ \times d_{5/2})$, and channels $(2^+ \times d_{5/2})$, $(2^+ \times s_{1/2})$, respectively. Panels (a) and (b) show the real and imaginary parts of the diagonal equivalent potential $V_{00}^{(eq)}$ in the entrance channel. Panels (c), (d) and (e), (f) depict the off-diagonal equivalent local potentials $V_{01}^{(eq)}$ and $V_{02}^{(eq)}$, respectively. Panels (g), (h) show the diagonal equivalent potential $V_{22}^{(eq)}$. The dotted and solid lines represent the equivalent local potentials corresponding to the first iteration and to the final solution of the CC equations, respectively.}
\end{figure}

\begin{figure}[hbt]
\center
\includegraphics[width=0.5\textwidth]{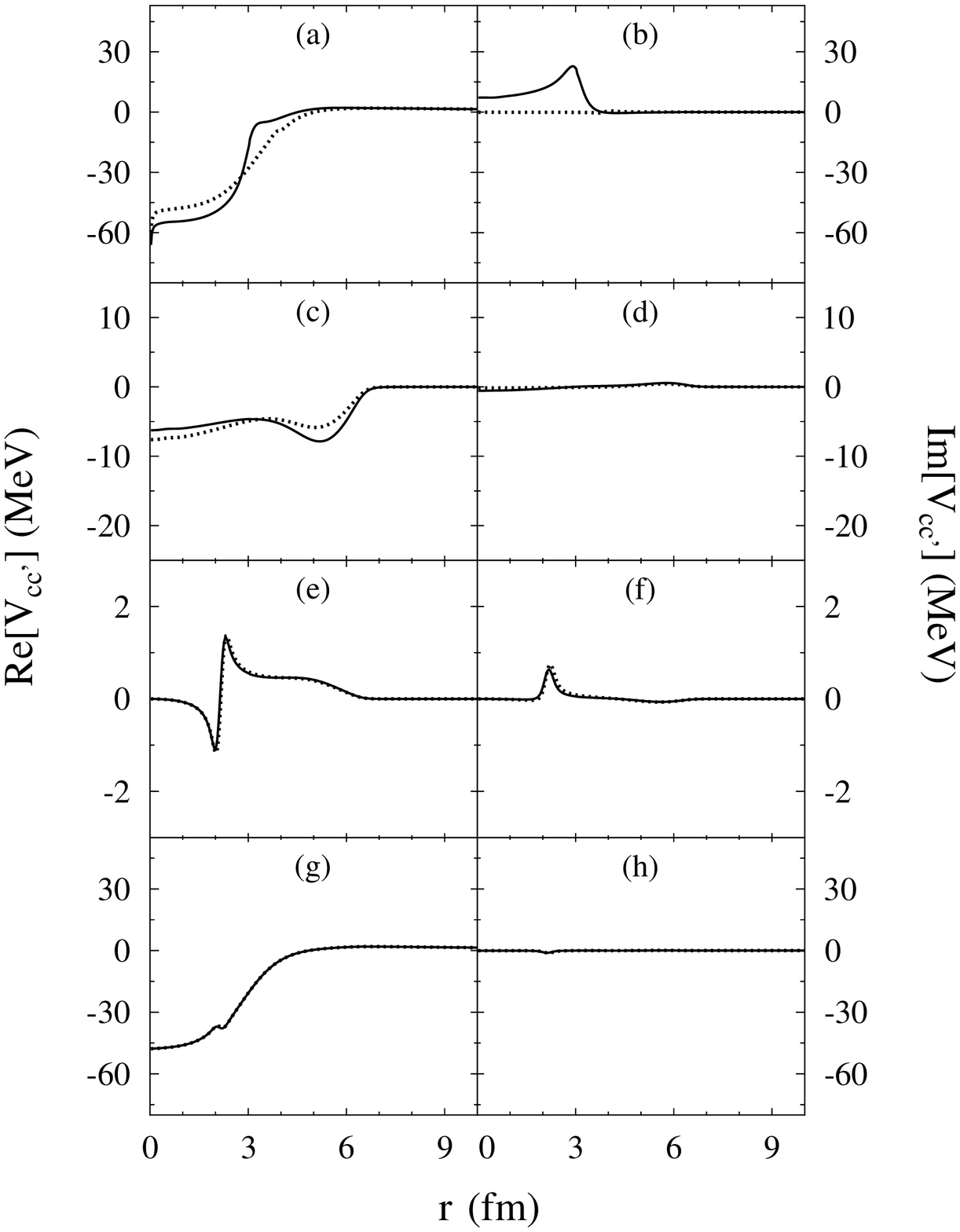}
\caption{\label{fig6} The same as in Fig. \ref{fig5} but for $E_{\rm CM}=$ 5 MeV.}
\end{figure}

\begin{figure}[hbt]
\center
\includegraphics[width=0.5\textwidth]{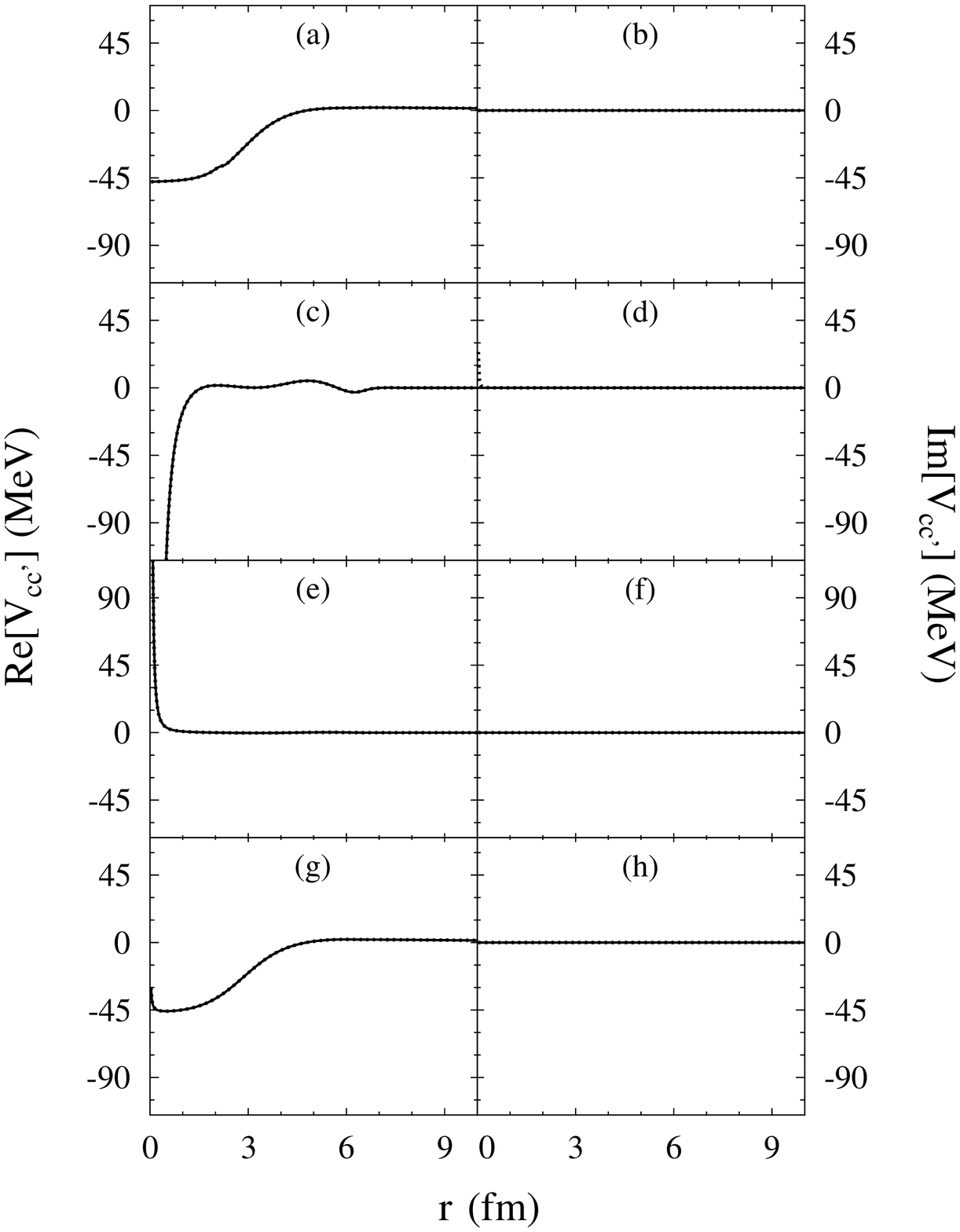}
\caption{\label{fig7} The equivalent local potentials $V_{cc'}^{(eq)}$ (Eq. (\ref{eq_pot_GSM})) for $J^\pi_A=1/2^+$ at $E_{\rm CM}=1$ MeV. The channel indices 0, 1, 2 denote the entrance channel $(0^+ \times s_{1/2})$, and channels $(2^+ \times d_{5/2})$, $(2^+ \times s_{1/2})$, respectively. Panels (a) and (b) show the real and imaginary parts of the diagonal equivalent potential $V_{00}^{(eq)}$ in the entrance channel. Panels (c), (d) and (e), (f) depict the off-diagonal equivalent local potentials $V_{01}^{(eq)}$ and $V_{02}^{(eq)}$, respectively. Panels (g), (h) show the diagonal equivalent potential $V_{22}^{(eq)}$. The dotted and solid lines represent the equivalent local potentials corresponding to the first iteration and to the final solution of the CC equations, respectively.
 }
\end{figure}

\begin{figure}[hbt]
\center
\includegraphics[width=0.5\textwidth]{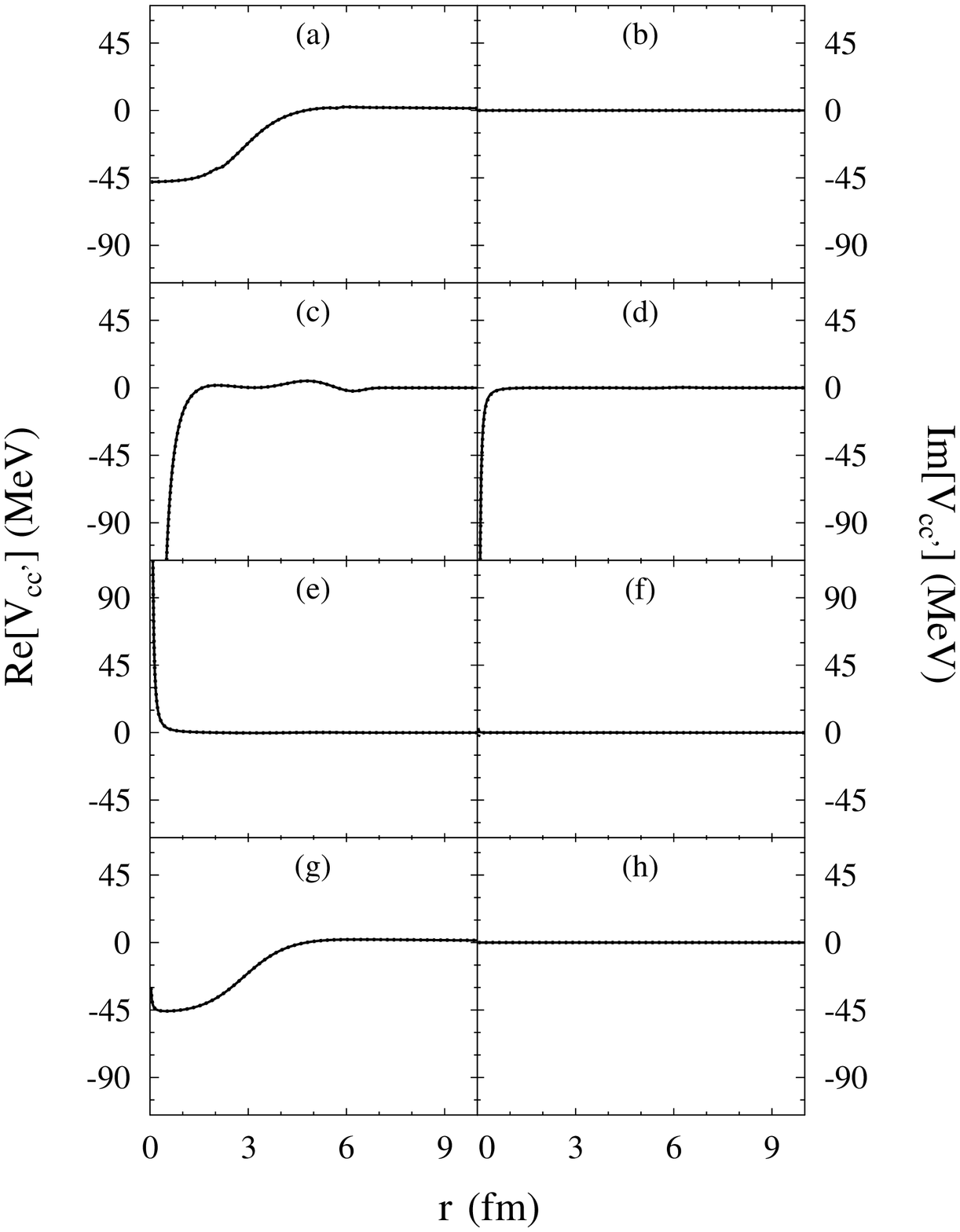}
\caption{\label{fig8} The same as in Fig. \ref{fig7} but for $E_{\rm CM}=$ 5 MeV.}
\end{figure}

Figs. \ref{fig5}-\ref{fig8} show the diagonal and off-diagonal parts of the equivalent local potentials for $J^\pi_A=5/2^+$ and $J^\pi_A=1/2^+$ at the two different CM energies $E_{\rm CM}=1$ MeV and 5 MeV. Figs. \ref{fig5} and \ref{fig6} exhibit results for $J^\pi_A=5/2^+$ at $E_{\rm CM}=$ 1 MeV and 5 MeV, respectively. The dotted and solid lines show the equivalent potentials at the first iteration and those corresponding to the final solution of the CC equations, respectively. For the detailed description of potential in each panel, see the caption of Fig. \ref{fig5}. One may notice that initial and final potentials are generally rather close. Small deviations can be seen mainly for the diagonal potential $V_{00}^{(eq)}$ which in this case corresponds to the entrance channel $(0^+ \times d_{5/2})$. 



Fig. \ref{fig7} and \ref{fig8} exhibit results for $J^\pi_A=1/2^+$ at $E_{\rm CM}=$ 1 MeV and 5 MeV, respectively. In this particular case, the initial and final equivalent local potentials are almost identical.

\subsubsection{$p+^{18}$Ne excitation function}

\begin{figure}[hbt]
\center
\includegraphics[width=0.5\textwidth]{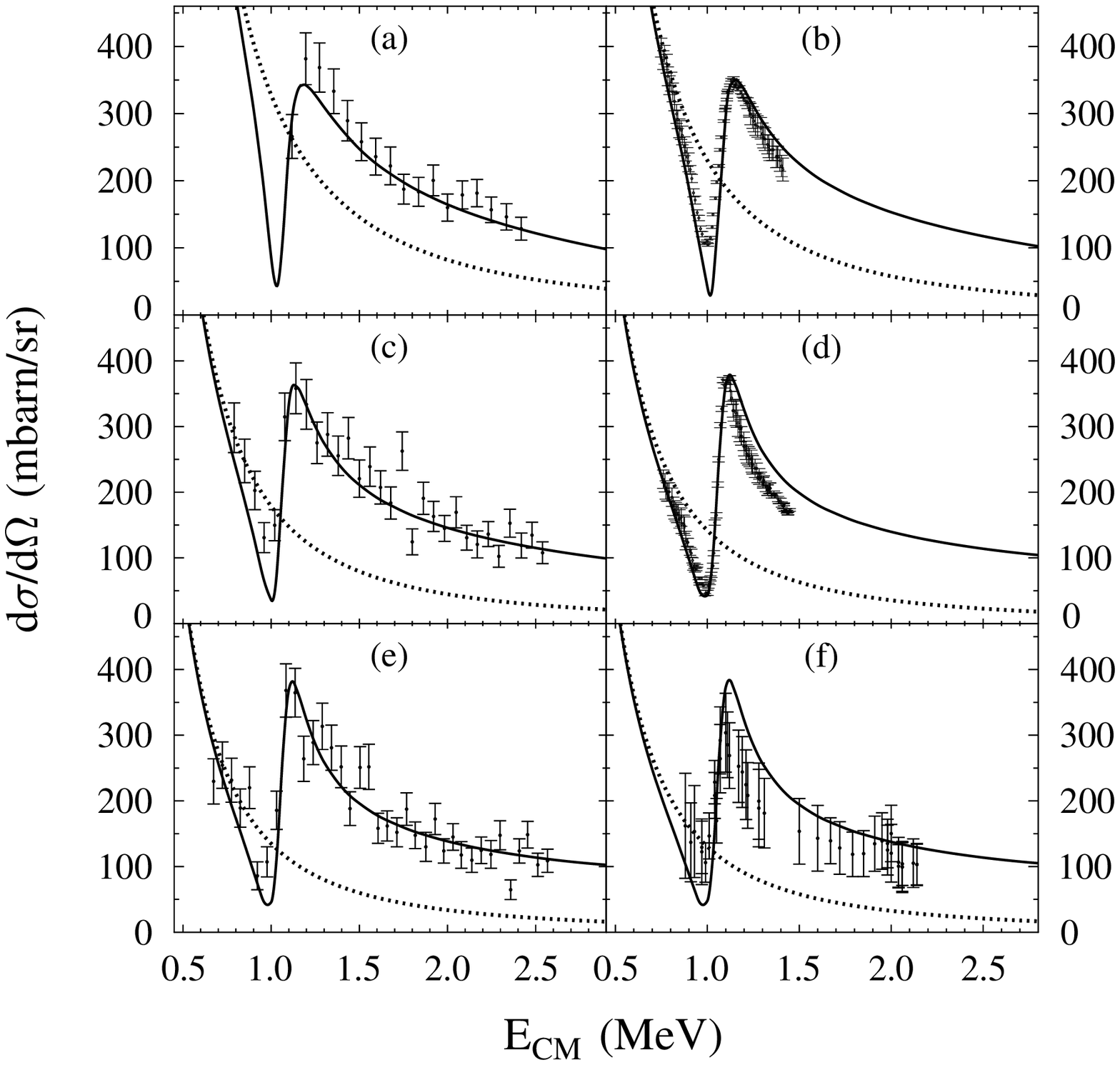}
\caption{\label{figEF} $p$+$^{18}$Ne excitation function at different CM angles $\Theta_{CM}$ in the  range from 105 to 180 deg. GSM-CC results are drawn with the solid line. The dotted line corresponds to the GSM-CC calculation using only Coulomb interaction. Panels (a), (b), (c), (d), (e), and (f) correspond to the excitation functions at $\Theta_{CM}$ = 105, 120.2, 135, 156.6, 165, and 180 deg, respectively. The experimental data at $\Theta_{CM}$ = 105, 135, and 165 deg are taken from Ref. \cite{sko06}. The data at $\Theta_{CM}$ = 120.2, 156.6 deg are taken from Ref. \cite{ang03} and at $\Theta_{CM}$ = 180 deg from Ref. \cite{oli05}. }
\end{figure}

The low-energy excitation function for the reaction $p+^{18}$Ne at different CM angles is plotted in Fig. \ref{figEF}. The solid line shows the GSM-CC excitation functions calculated using the full Hamiltonian which includes both nuclear and Coulomb interactions. The dotted line corresponds to the GSM-CC calculation with the Coulomb interaction only. The peak in the excitation function at $\sim$1.1 MeV corresponds to the $1/2^+$ resonance (see the Table \ref{tab:3}).
Experimental data are also represented and the GSM-CC calculation are in very good accordance at all angles. 

\subsubsection{Elastic and inelastic $^{18}$Ne(p,p') reaction cross-sections}

\begin{figure}[hbt]
\center
\includegraphics[width=0.5\textwidth]{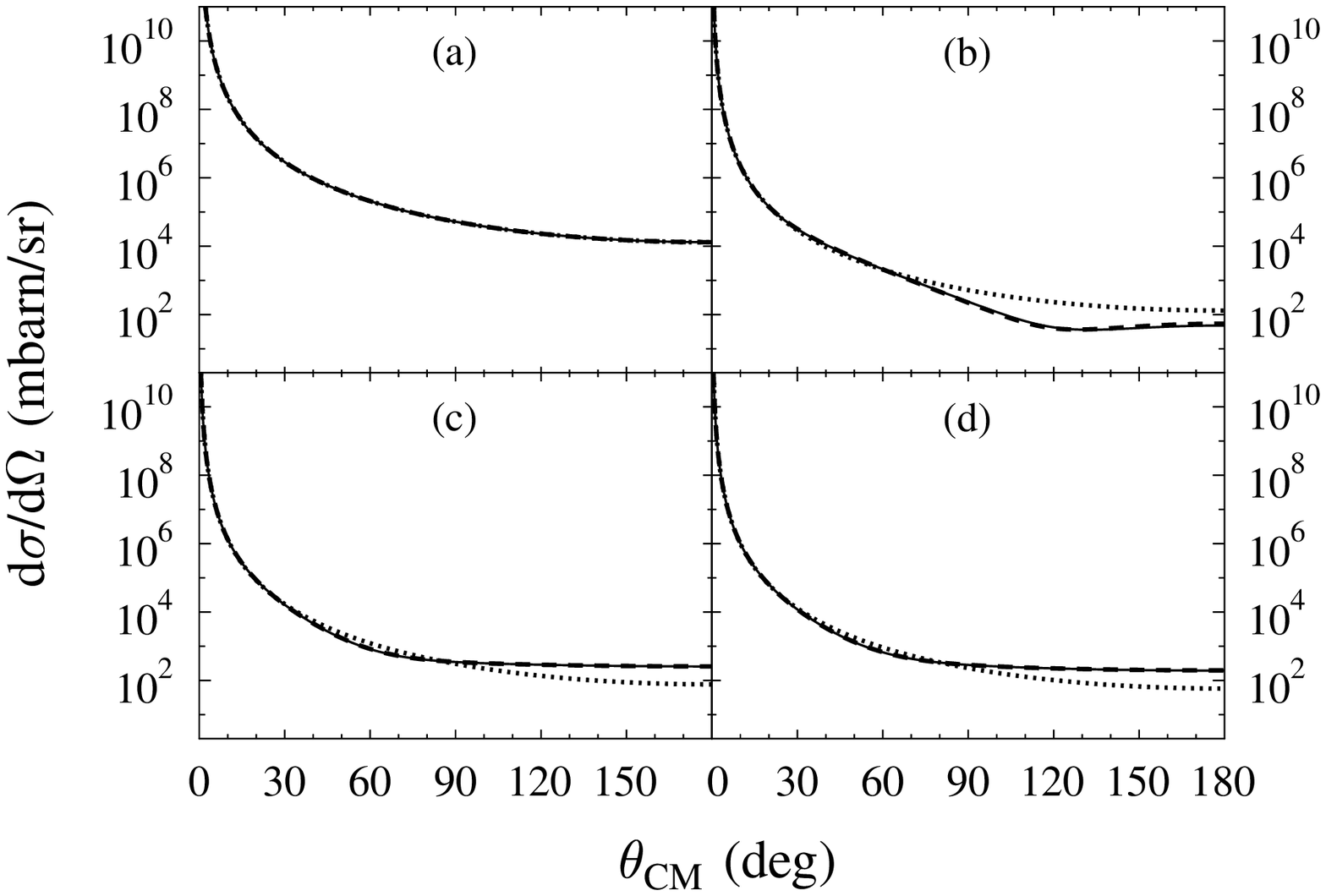}
\caption{\label{figBT} Elastic angular cross sections for the reaction $p+^{18}$Ne at different CM energies:  $E_{CM}$ = 0.1 MeV (panel (a)), 1.0 MeV (panel (b)), 1.3 MeV (panel (c)) et 1.5 MeV (panel (d)) below the inelastic channel threshold. Solid and dashed lines show the GSM-CC results in $(sd-p)$ and $(sd)$ model spaces, respectively. Dotted lines have been obtained using the Coulomb interaction only.}
\end{figure}

\begin{figure}[hbt]
\center
\includegraphics[width=0.5\textwidth]{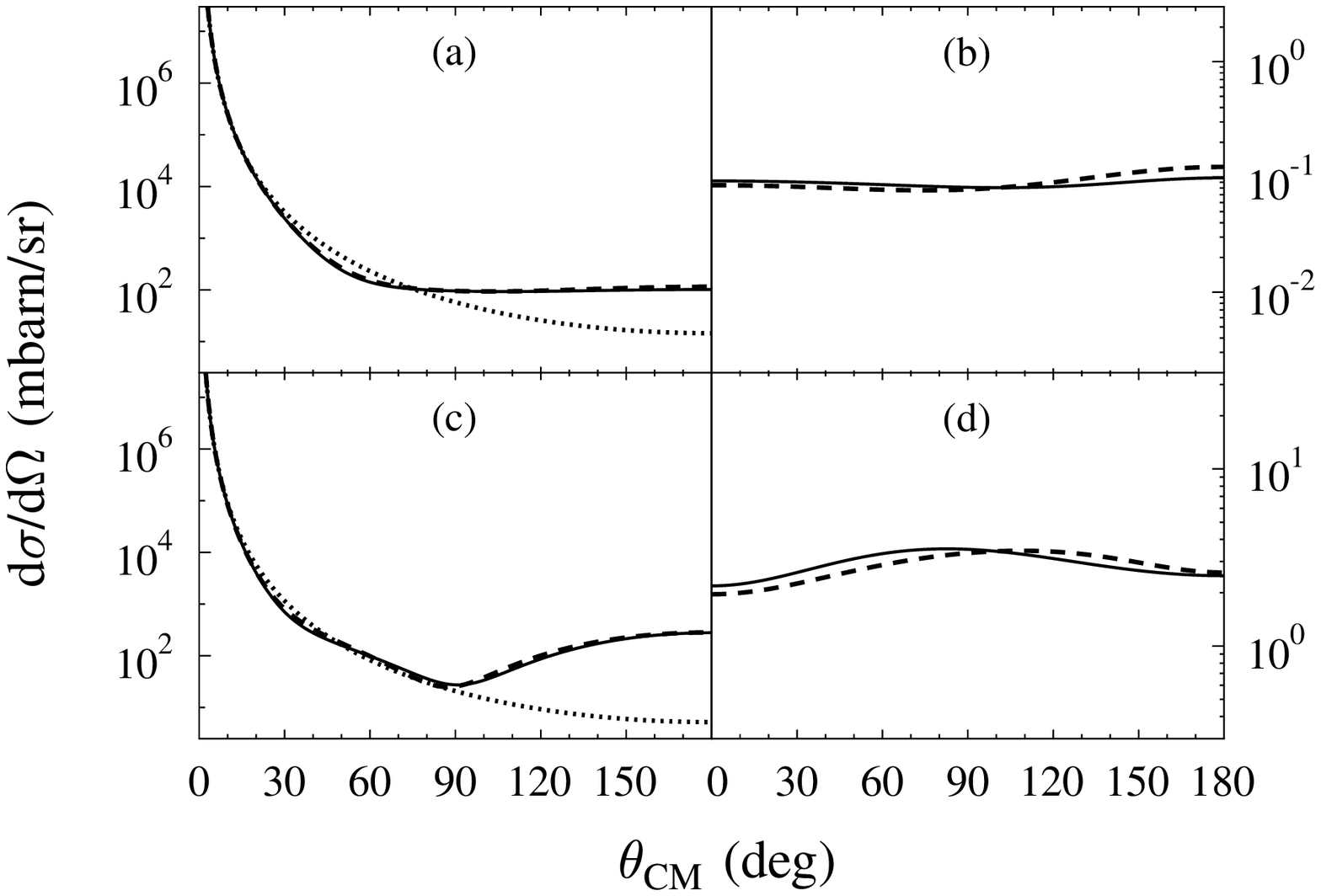}
\caption{\label{figAT} Elastic (left) and inelastic (right) angular cross sections for the reaction $p+^{18}$Ne at different CM energies $E_{CM}$~=~3 MeV (panels (a) and (b)) and 5 MeV (panels ((c) and (d)) above the inelastic channel threshold. Solid and dashed lines show the GSM-CC results in $(sd-p)$ and $(sd)$ model spaces, respectively. Dotted lines show results obtained using the Coulomb interaction only.}
\end{figure}

Figs. \ref{figBT} and \ref{figAT} show the $p+^{18}$Ne angular cross-sections as a function of the angle $\Theta_{CM}$ for the excitation energies in the range from  $E_{CM}$ = 0.1 MeV to 5 MeV.  Fig. \ref{figBT} represents the GSM-CC results below the inelastic channel threshold at 1.887 MeV. Panels (a), (b), (c), and (d) in Fig. \ref{figBT} present elastic cross-sections at $E_{CM}$ = 0.1, 1.0, 1.3, and 1.5 MeV, respectively. 

Solid and dashed lines show the GSM-CC cross-sections which are calculated in $(sd-p)$ and $(sd)$ model spaces, respectively. These two calculations give indistinguishable results. The dotted lines have been obtained neglecting nuclear interaction and using the Coulomb interaction only. The comparison of solid and dotted lines allows to assess the role of the nuclear interaction in the calculated angular cross-sections. 

Results shown in Fig. \ref{figAT} have been obtained for energies $E_{CM}$ = 3 and 5 MeV, above the inelastic channel threshold. GSM-CC angular cross-sections calculated in $(sd-p)$ (solid lines) and $(sd)$ (dashed lines) model spaces are similar. Deviations from the pure Coulomb scattering (dotted lines) is well seen in the elastic cross-sections at backward angles.

\section{Conclusions\label{sec5}}

In this work, we have presented in details  the GSM in the coupled channel representation and applied it for the description of elastic and inelastic scattering of protons on a heavy target which is described by a limited set of GSM states. By combining the new method of solving integro-differential CC equation based on the equivalent potential method with the GSM algorithm to diagonalize the Hamiltonian matrix in the Berggren basis, we were able to perform the many-body calculation for the reaction $^{18}$Ne$(p,p')$, and determine ground state and excited states of the unbound nucleus $^{19}$Na. The interaction between valence nucleons in this calculation was modelled by the finite-range MSG interaction. 
 
 The convergence of GSM-CC calculations has been carefully checked by comparing GSM and GSM-CC results for $^{19}$Na resonances. In a given s.p. model space, the GSM-CC calculation with the reaction channels which are constructed using selected many-body states of the target, can be considered reliable if the GSM-CC eigenvalues for an intermediate system, in our case $^{19}$Na, approximate well the results of a direct diagonalization of the GSM Hamiltonian matrix in the same s.p. model space. In such a case, the configuration mixing in GSM-CC and GSM wave functions are equivalent and one does not need to include additional states of the target nucleus to reach the many-body completeness in GSM-CC calculation. Only in this case, the unified description of nuclear structure and reactions with the same many-body Hamiltonian and the same model space is possible. This situation has been achieved in this work for low-energy proton scattering on well-bound nucleus $^{18}$Ne ($S_p=3.921$ MeV, $S_n=19.237$ MeV) with proton-unbound intermediate nucleus $^{19}$Na ($S_p=-0.323$ MeV, $S_n=20.18$ MeV), considering the ground state $0_1^+$ and the first excited state $2_1^+$ of $^{18}$Ne in the construction of reaction channels. 
  
 Numerical tests of the stability of GSM-CC calculations have been discussed on the examples of wave functions and effective local potentials which were calculated at different CM energies. Excellent convergence has been achieved if the initial point of the GSM-CC iterative calculation is given by the diagonalization of the modified Hamiltonian $\mathcal{H}_m$ (Eq. (\ref{Hm})) in Berggren basis.

In the near future, we plan to apply the GSM-CC formalism for the description of proton/neutron scattering on weakly bound targets, such as $^6$He. In this case, it is expected that to achieve a unified description of structure and reaction in the GSM framework, a large number of target states has to be taken to construct the reactions channels. Further developments of the GSM-CC formalism to describe proton radiative capture reactions and $(d,p)$, $(p,d)$ transfer reactions are in progress.


\bigskip
\begin{acknowledgments}
N. Michel is grateful for a support extended to him by the Oak Ridge National Laboratory, University of Tennessee, and Michigan State University were part of the results have been obtained.  This work is a part of the PhD thesis of Y. Jaganathen prepared at GANIL. We thank W Nazarewicz and F. Nunes for useful discussions. Part of the computations have been performed on Kraken at the National Institute for Computational Sciences (http://www.nics.tennessee.edu/) and on Titan at the Oak Ridge Leadership Computing Facility at the Oak Ridge National Laboratory, which is supported by the Office of Science of the U.S. Department of Energy. This work has been supported in part by the by U.S. Department of Energy under Contract Nos. DE-FG02-96ER40963 (University of Tennessee) and DE-FG02-10ER41700 (French-U.S. Theory Institute for Physics with Exotic Nuclei).
\end{acknowledgments}


\end{document}